\documentclass[11pt]{article}

\usepackage{graphics}
\usepackage{subfigure}

\newcommand{\beq}{\begin{equation}}
\newcommand{\eeq}{\end{equation}}

\newcommand{\remove}[1]{}

\newcommand{\lae}{\stackrel{<}{\sim}}
\newcommand{\gae}{\stackrel{>}{\sim}}

\title{Current Bounds on Technicolor with Scalars}

\author{Vagish Hemmige\thanks{hemmige@fas.harvard.edu}$^{\ 2}$ and 
  Elizabeth H. Simmons\thanks{simmons@bu.edu}$^{\ 1, 2}$\\
  \\
  $^1$ Department of Physics, Boston University, \\
  590 Commonwealth Avenue, Boston MA  02215\\
  \\
  $^2$ Radcliffe Institute for Advanced Study and\\
 Jefferson Laboratory of Physics \\
 Harvard University, Cambridge, MA  02138}

\begin{document}

\begin{titlepage}
\maketitle
\thispagestyle{empty}

  \begin{picture}(0,0)(0,0)
    \put(400,300){BUHEP-01-10}
    \put(400,290){HUPT-01/A033}
  \end{picture}
  \vspace{24pt}


\begin{abstract}
  Technicolor with scalars is the simplest dynamical symmetry breaking model
  and one in which the predicted values of many observables may be readily
  calculated.  This letter applies current LEP, 
  Tevatron, CESR, and SLAC data from searches for neutral and charged scalars
  and from studies of $b$ physics to obtain bounds on technicolor with
  scalars.  Expectations for how upcoming measurements will further probe the
  theory's parameter space are also discussed.

\end{abstract}

\end{titlepage}

\newpage
\renewcommand{\thepage}{\arabic{page}}
\setcounter{page}{1}

\section {Introduction} \label {sec:intro}

Technicolor theories\cite{technicolor} can successfully break the electroweak
symmetry, but require additional interactions to communicate the symmetry
breaking to the quarks and leptons.  In extended technicolor
theories\cite{extendedtc}, the additional interactions are gauge
interactions; arranging for gauge bosons to generate the wide range of
observed fermion masses without causing excessive flavor-changing neutral
currents\cite{extendedtc}, large weak isospin violation\cite{app-rho}, or
contributions to other precision electroweak observables\cite{prewrc,
  Chivukula:1992ap} is tricky.  An alternative is to consider a low-energy
effective theory in which the additional fields that connect the technicolor
condensate to the ordinary fermions are scalars \cite{sim}.  Such scalars
can, for example, arise as composite bound states in strongly-coupled
extended technicolor theories \cite{setc}, have masses protected by
supersymmetry\cite{sdk, Dobrescu:1995gz}  
or be associated with TeV-scale extra
dimensions \cite{extradim}.

This paper assesses current experimental constraints on technicolor models
with scalars.  The phenomenology of these models has been considered
extensively in the literature \cite{Chivukula:1992ap, sim,
  sdk,  evans, caronet, Carone:1995mx,
  yumian, xiong}.  It has been found that these theories do not produce
unacceptably large contributions to neutral meson mixing, or to the
electroweak $S$ and $T$ parameters\cite{sim,caronet}.  Indeed, the effect of
the weak-doublet scalar on the electroweak vacuum alignment renders viable an
$SU(2)$ technicolor group, with its attendant small oblique corrections
\cite{Dobrescu:1999ci}.  On the other hand, the models do predict potentially
visible contributions to b-physics observables such as $R_b$
\cite{Carone:1995mx} and the rate of various rare $B$-meson decays
\cite{Carone:1995mx, yumian, xiong}.

In section 2, we review the minimal model, focusing on information
relevant to comparing theory with experiment. Section 3 explores the
constraints imposed by searches for neutral and charged scalar bosons,
by measurements of $R_b$, and by other heavy flavor observables.  We
also indicate how upcoming measurements will further probe the
theory's parameter space.  Section 4 discusses our conclusions.

\section{The Model}

The theory\footnote{For a more detailed description, see \cite{sim, caronet}.}
includes the full Standard Model gauge structure and fermion content; all of
these fields are technicolor singlets.  There is also a minimal
$SU(N)$ technicolor sector, with two techniflavors that transform as a
left-handed doublet and two right-handed singlets under $SU(2)_W$, 
\beq
T_L=\left(\begin{array}{c} p \\ m
\end{array} \right)_L  \,\,\,\,\,p_R \,\,\,\, m_R
\eeq
with weak hypercharges $Y(T _L)=0$,
$Y(p_R)=1/2$, and $Y(m_R)=-1/2$.  All of the fermions couple to a weak scalar
doublet which has the quantum 
numbers of the Standard Model's Higgs doublet 
\beq
\phi=\left(\begin{array}{c} \phi^+ \\ \phi^0
\end{array}\right)
\eeq 
This scalar's purpose is to couple the technifermion condensate to the
ordinary fermions and thereby generate fermion masses.  It has a non-negative
mass-squared and does not trigger electroweak symmetry breaking.  However,
when the technifermions condense (with technipion decay constant $f$), their
coupling to $\phi$ induces a vacuum expectation value (vev) $f'$.  Both the
technicolor scale and the induced vev contribute to the electroweak scale $v
= 246$ GeV: 
\beq f^2+f'^2=v^2
\label{ews}
\eeq

Our analysis depends on the properties of
the scalars left in the spectrum after spontaneous electroweak
symmetry breaking.  The technipions (the isotriplet scalar bound states of
$p$ and $m$) and the isotriplet components of $\phi$ will mix.  One linear
combination becomes the longitudinal component of the $W$ and $Z$.  The
orthogonal linear combination (which we call $\pi_p$) remains in the
low-energy theory as an isotriplet of physical scalars.  In addition, the
spectrum contains a ``Higgs field'': the isoscalar component of the $\phi$
field, which we denote $\sigma$.

The coupling of the
charged physical scalars to the quarks is given by \cite{caronet}
\beq
i(\frac{f}{v})\left[ \overline{D_L} V^\dagger \pi^-_p h_U U_R
+\overline{U_L} \pi^+_p V h_D D_R + h.c.\right]
\label{cpcoup}
\eeq
where $V$ is the Cabibbo-Kobayashi-Maskawa (CKM) matrix, $U$ and $D$ are
column vectors of ordinary quarks in flavor space, and the Yukawa coupling
matrices are diagonal $h_U=diag(h_u,h_c,h_t)$, $h_D=diag(h_d,h_s,h_b)$.
Notice that (\ref{cpcoup}) has the same form as the charged scalar coupling
in a type-I two-Higgs doublet model; the dependence of (\ref{cpcoup}) on
$f/v$ arises because the quarks couple to $\phi$ and not to the technipions.

A chiral Lagrangian analysis \cite{caronet} of the theory below the
symmetry-breaking scale estimates the masses of the $\pi_p$ to be
\beq
m_{\pi_p}^2=2c_1\sqrt{2}\frac{4\pi f}{f'} v^2 h
\label{mpip}
\eeq
where $h$ is the average technifermion Yukawa coupling
$h\equiv (h_++h_-)/2$, and where $h_+$ and $h_-$ are the individual Yukawa
couplings to $p$ and $m$, respectively.  The constant $c_1$ is an
undetermined coefficient in the chiral expansion, but is of order unity by
naive dimensional analysis (NDA) \cite{nda}.  We set $c_1=1$ from here on.
 As we work to lowest order,
$c_1$ and $h$ always appear in the combination $c_1 h$; the uncertainty
in $c_1$ can, thus, be expressed as an uncertainty in the value of $h$.

The behavior of $\sigma$ is governed by its effective potential,
which at one loop has the 
form~\cite{caronet},
\begin{equation}\label{potential}
V(\sigma)=\frac{1}{2}{M_\phi}^2\sigma^2 +\frac{\lambda}{8}\sigma^4 
-\frac{1}{64\pi^2}\left[3h_t^4
+N(h_+^4+h_-^4))\right]\sigma^4\log{\left(\frac{\sigma^2}{\mu^2}\right)}
-8\sqrt{2}c_1\pi f^3h\sigma,
\end{equation}
where $h_t$ is the top quark Yukawa coupling ($h_t=\sqrt{2}m_t/f^\prime$),
$N=4$, and $\mu$ is an arbitrary renormalization scale. The first three 
terms in equation (\ref{potential}) are standard one loop 
Coleman--Weinberg terms~\cite{Coleman:Weinberg}.
The last term enters through the technicolor interactions. 

Technicolor plus scalars requires four parameters, beyond those of the
Standard Model, to fully specify the theory:
$(M_\phi, \lambda, h_+, h_-)$.  The literature 
studies two limits of the model: {\it [i]} the limit in which $\lambda$ is
negligibly small; and {\it [ii]} the limit in which $M_\phi$ is negligibly
small.  

\subsection{Limit [i]:  $\lambda\approx 0 $}

Because the scalar $\phi$ does not trigger electroweak symmetry breaking,
the $\sigma$ field has no vev and 
terms in the potential $V(\sigma)$ that are linear in $\sigma$ should vanish:
\begin{equation}
\label{conditionn}
 V^\prime(\sigma)=0. 
\end{equation}
Applying this to equation (\ref{potential}) in the limit where the $\phi^4$
coupling 
vanishes gives the relation
\begin{equation}\label{no:VEV:1}
{\widetilde{M}_\phi}^2f^\prime=8\sqrt{2}c_1\pi hf^3,
\end{equation}
where the shifted scalar mass $\widetilde{M}_\phi$ is connected to the
unshifted mass $M_\phi$ by the Coleman-Weinberg corrections 
\begin{equation}\label{shifted:mass}
{\widetilde{M}_\phi}^2=M_\phi^2
+\left(\frac{44}{3}\right)\frac{1}{64\pi^2}\left[3h_t^4
+2Nh^4\right]{f^\prime}^2.
\end{equation}
In deriving equations (\ref{no:VEV:1}) and (\ref{shifted:mass}), we have
defined the  
renormalized $(\phi^{\dag}\phi)^2$ coupling as
$\lambda_r=V^{\prime\prime\prime\prime}(f^\prime)/3$ to remove the $\mu$
dependence. For simplicity, we also set $h_+=h_-$ in
eq. (\ref{shifted:mass}).  By using the shifted scalar mass, we can
absorb radiative corrections which affect the phenomenology of the
charged scalar. However, these corrections still appear in the mass of
the $\sigma$ field, which is determined by $V^{\prime\prime}(f^\prime)$ to
be: 

\begin{equation}
 { m_\sigma}^2= \widetilde{M}_\phi^2+
\left(\frac{64}{3}\right)\left(\frac{1}{64\pi^2}\right)\left[3h_t^4
+2Nh^4\right]{f^\prime}^2.
\end{equation}

In this limit, the phenomenology can be described in terms of
$(\widetilde{M}_\phi,h)$, as has been done in some of the literature
\cite{sim}, \cite{caronet}-\cite{yumian}.  Alternatively, we can trade the
unphysical parameter $\widetilde{M}_\phi$ for the mass of the isoscalar
field, $m_\sigma$, as in refs. \cite{yumian, xiong}. Then the free parameters
will be two physical quantities: $(m_\sigma, h)$.

\subsection{Limit [ii]: $M_\phi\approx 0$}

Applying condition (\ref{conditionn}) to the effective potential
(\ref{potential}) in limit [ii] yields the relation 
\begin{equation}\label{no:VEV:2}
\frac{\tilde{\lambda}}{2}{f^\prime}^3=8\sqrt{2}c_1\pi hf^3,
\end{equation}
where the shifted coupling $\tilde{\lambda}$ is defined by

\begin{equation}
 \tilde{\lambda}=\lambda+\frac{11}{24\pi^2}\left[3h_t^4 + 2Nh^4\right].
\end{equation}
The same renormalization scheme as that in limit {\it  [i]} is used.
The effects of radiative corrections are absorbed 
into the shifted coupling $\tilde{\lambda}$ but still manifest in the
$\sigma$ mass, which is given by 

\begin{equation}
{m_\sigma}^2=\frac{3}{2}\tilde{\lambda}{f^\prime}^2 
    - \frac{1}{8\pi^2}\left[3h_t^4+2Nh^4\right]{f^\prime}^2.
\end{equation}
In this limit, we can choose $(\tilde{\lambda},h)$ to be our free parameters
as in refs. \cite{caronet}-\cite{yumian} or use $(m_\sigma,h$) as in \cite{yumian, xiong}.

To the extent that these results depend on the effective chiral Lagrangian
analysis, they are valid only if the technifermion masses ($\approx
hf^\prime$) lie below the technicolor scale ($\approx 4\pi f$).  We will
see that this requirement is consistent with the experimentally allowed
region in limit [i] and that the experimental constraints always enforce $h
f^\prime < 4 \pi f$ in limit [ii].

\section{Results}

\begin{figure}[tb]
\begin{center}
  \begin{minipage}{3in}
    \begin{center}
      \rotatebox{90}{\scalebox{.35}{\includegraphics{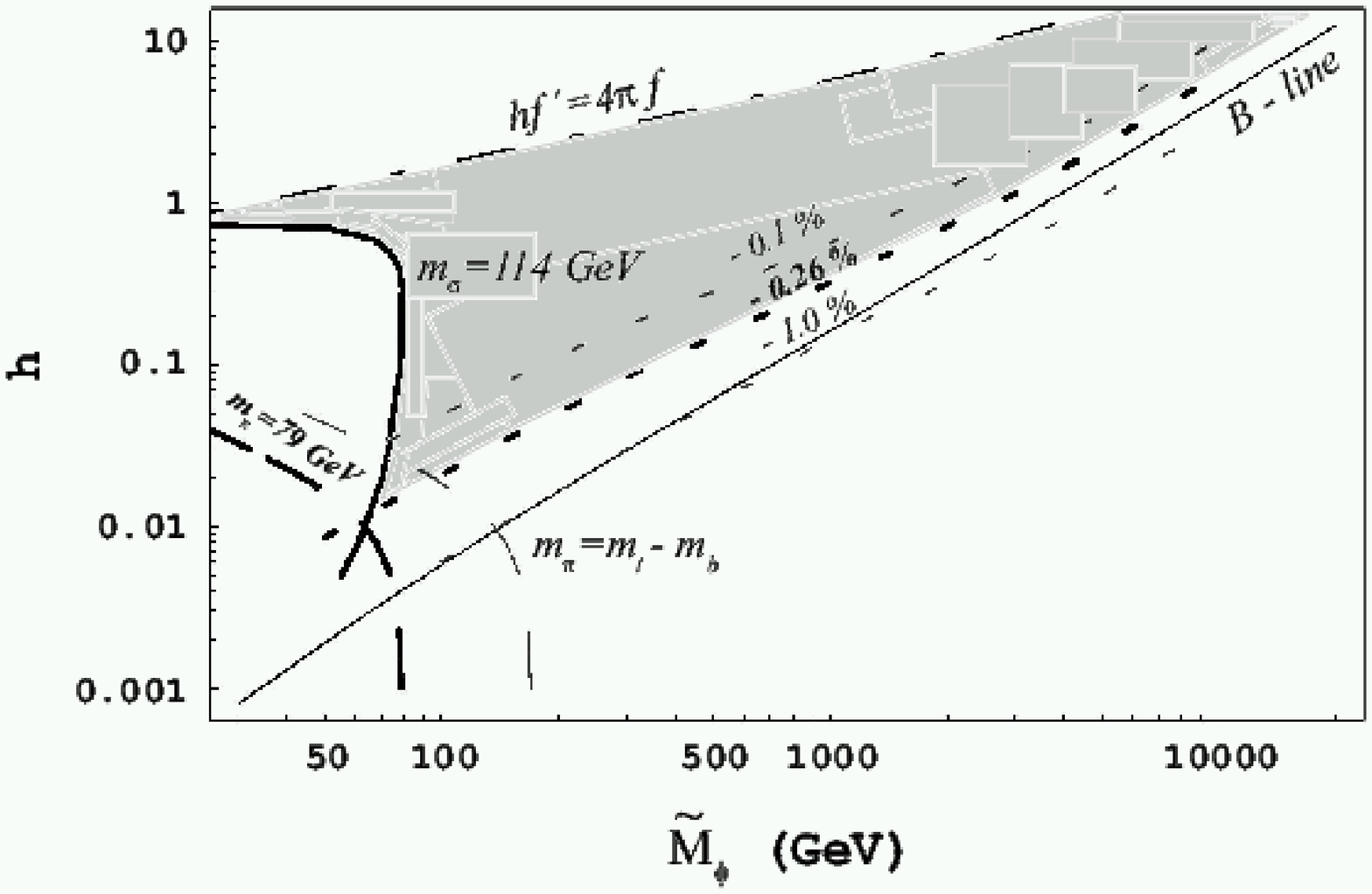}}}
    \end{center}
  \end{minipage}\qquad
  \begin{minipage}{3in}
    \begin{center}
      \rotatebox{90}{\scalebox{.35}{\includegraphics{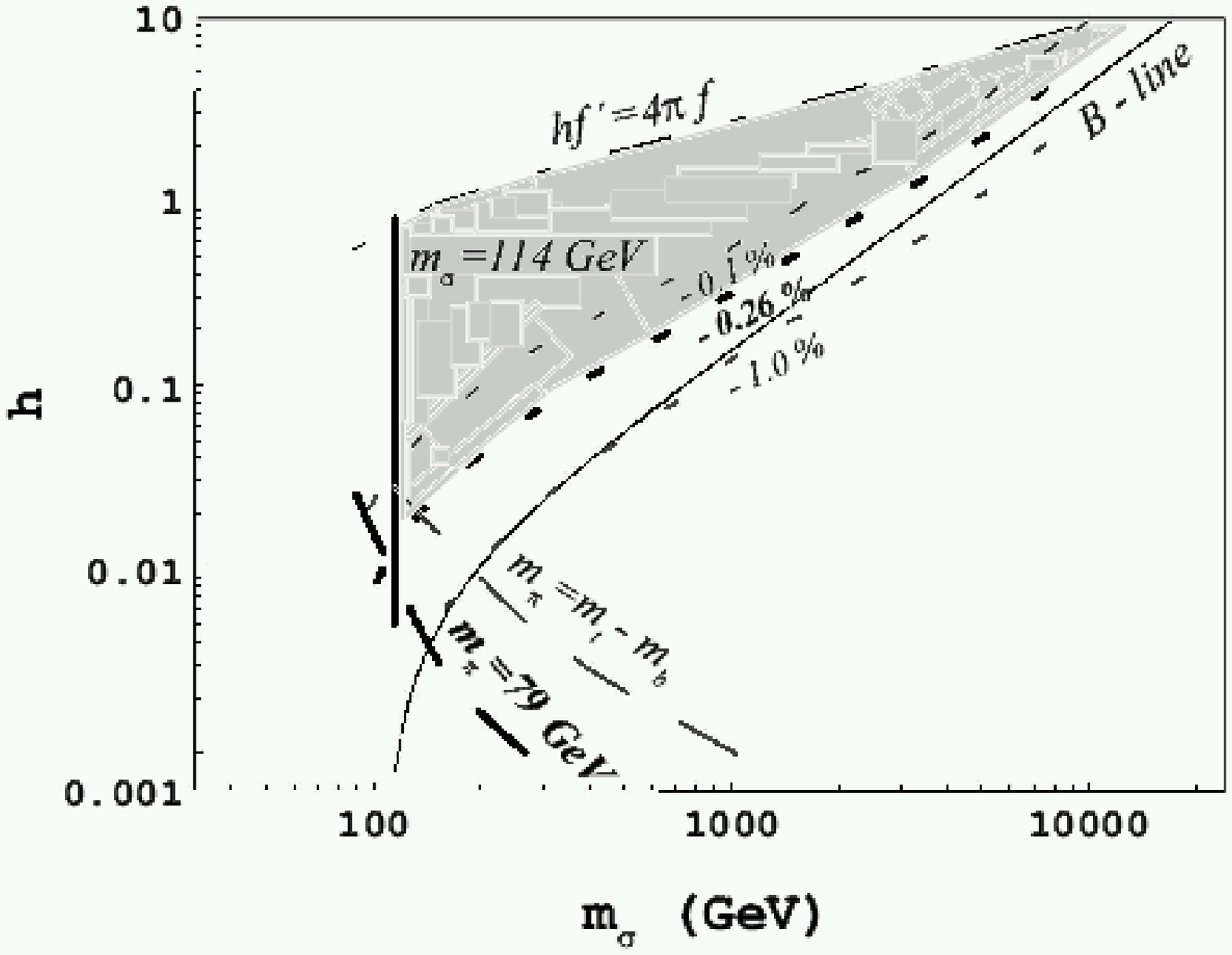}}}
    \end{center}
  \end{minipage}
\end{center}
\vspace{0cm}
\caption[limi]{\small Constraints on technicolor with scalars in limit [i],
where the scalar self-coupling is negligible, plotted on the left in the
conventional basis ($\tilde{M}_\phi$, h) and on the right in the
physical basis ($m_\sigma$, h). The allowed region of parameter space (shaded)
is bounded by the contours $m_\sigma = 114$ GeV (solid), $R_b -
R_b^{SM} = 0.26\%$ (dashes) and $h f' = 4 \pi f$ (dot-dash).  Other
contours of constant $R_b$ are shown for reference.
The current bound from searches for charged scalars $m_{\pi^\pm_p} =
79$ GeV is shown (long dashes) along with the reference curve
$m_{\pi^\pm_p} = m_t - m_b$.  The constraint from
$B^0\bar{B}^0$ mixing is labeled ``B-line''.}
\label{fig-limi}
\end{figure}

\begin{figure}[tb]
\begin{center}
  \begin{minipage}{3in}
    \begin{center}
      \rotatebox{90}{\scalebox{.35}{\includegraphics{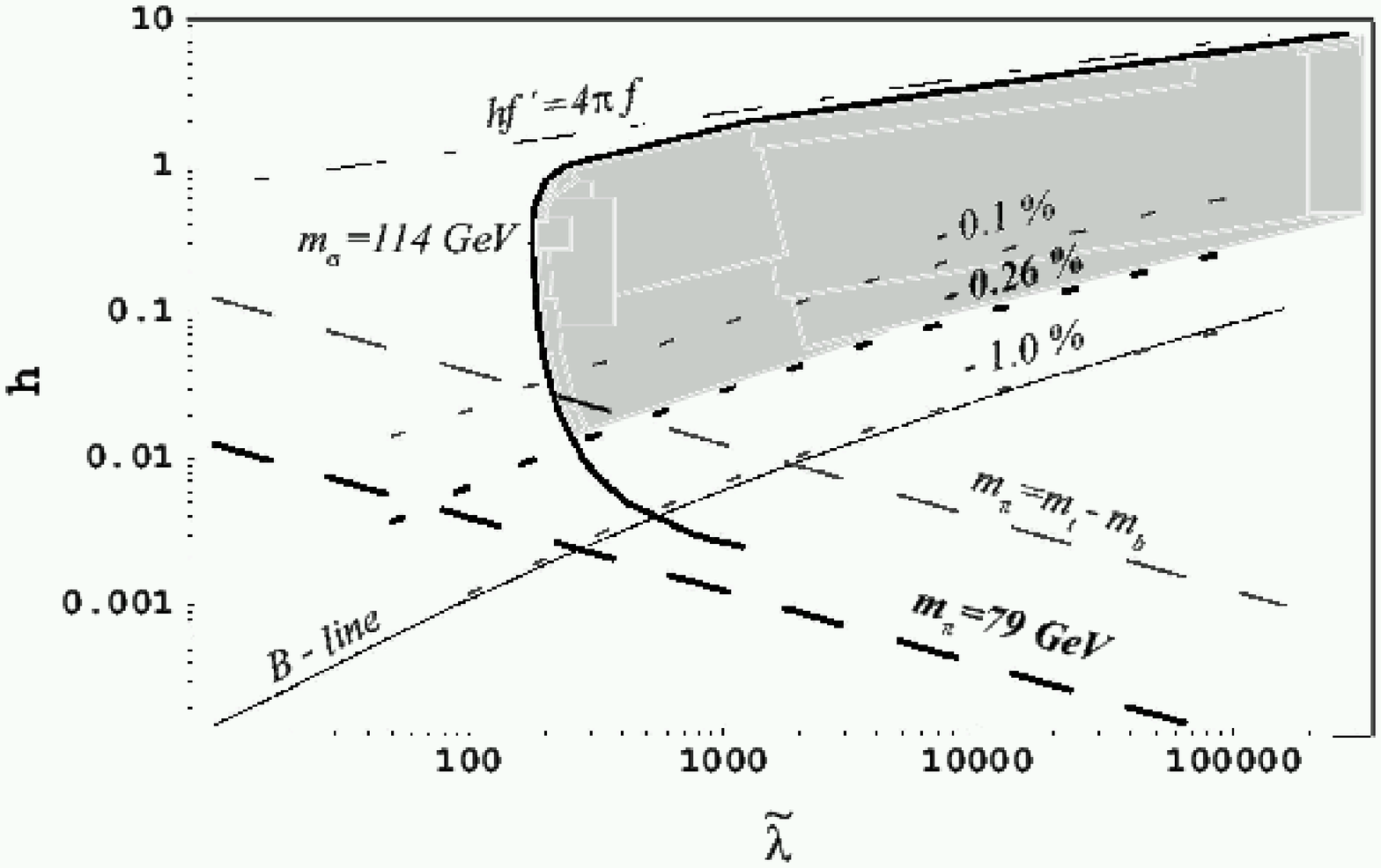}}}
    \end{center}
  \end{minipage}\qquad
  \begin{minipage}{3in}
    \begin{center}
      \rotatebox{90}{\scalebox{.35}{\includegraphics{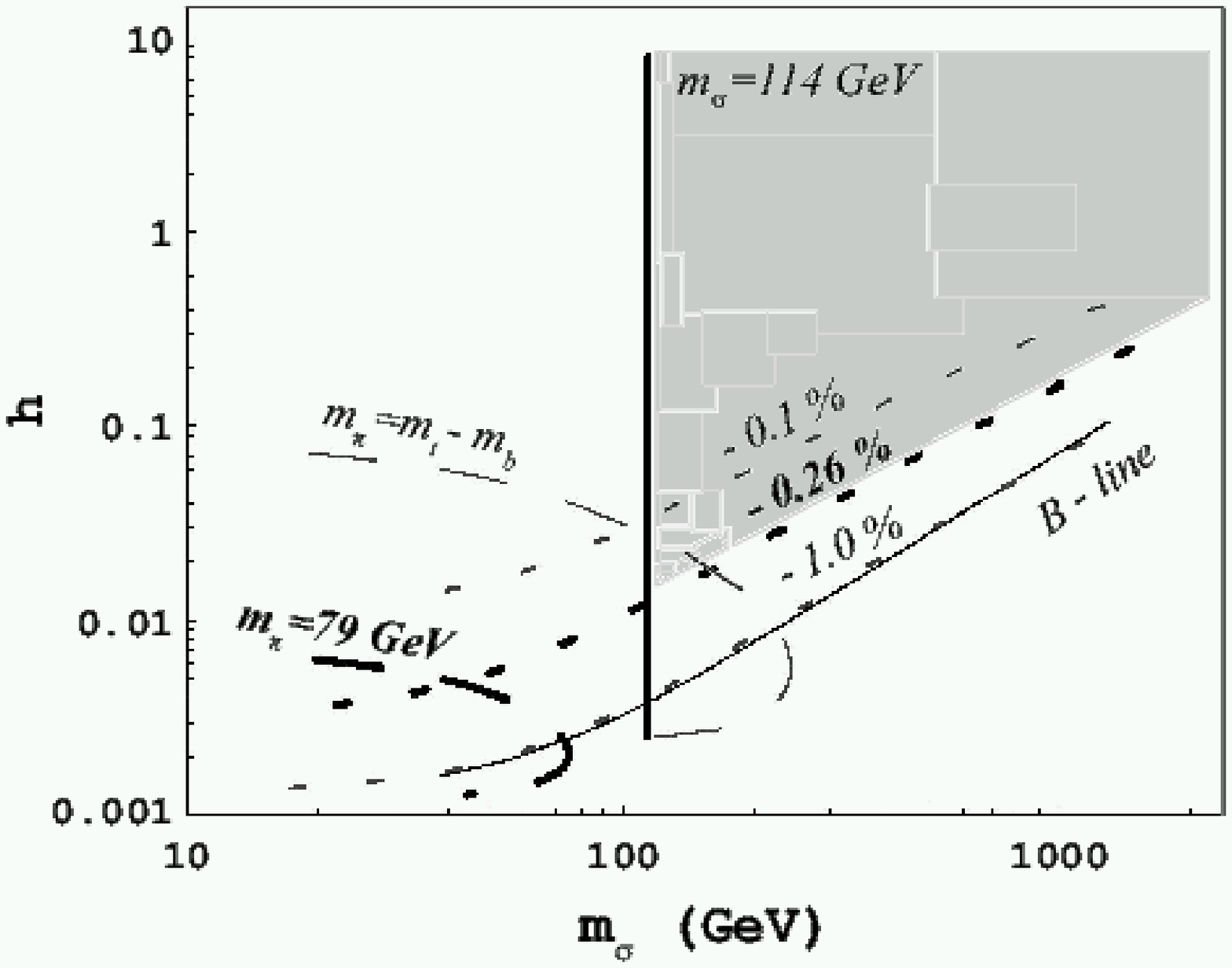}}}
    \end{center}
  \end{minipage}
\end{center}
\vspace{0cm}
\caption[limii]{\small Constraints on technicolor with scalars in limit
[ii], where the scalar mass is negligible, plotted on the left in the
conventional basis ($\tilde{M}_\phi$, h) and on the right in the
physical basis ($m_\sigma$, h). The allowed region of parameter space
is bounded by the contours $m_\sigma = 114$ GeV (solid) and $R_b -
R_b^{SM} = 0.26\%$ (dashes).  Other contours of constant $R_b$ are
shown for reference.  The current bound from searches
for charged scalars $m_{\pi^\pm_p} = 79$ GeV is shown (long dashes) along
with the reference curve $m_{\pi^\pm_p} = m_t - m_b$.
The constraint from $B^0\bar{B}^0$ mixing is
labeled ``B-line''; the theoretical constraint $h f' = 4 \pi f$
is also indicated.}
\label{fig-limii}
\end{figure}

We have assessed the current bounds on technicolor with scalars, using
data from a variety of sources.  Our results are summarized in Figures
\ref{fig-limi} and \ref{fig-limii}.  In each plot, the
allowed area is the shaded region.  Figure
\ref{fig-limi} is for limit [i], in which
$\lambda$ is assumed to be small; it shows the same information in
the conventional ($\tilde M_\phi$, h) and physical ($m_\sigma$, h)
parameterizations.  Likewise, Figure \ref{fig-limii} 
shows the results for limit [ii] in two formats.  We will now discuss
the origins and implications of the contours in the figures.

\subsection{$R_b$ and other $b$ physics}

Radiative corrections to hadronic $Z$ decays resulting from the presence of
the extra physical charged scalars in the low-energy spectrum tend to reduce
the value of $R_b$ below the Standard Model prediction in models of
technicolor with scalars.  The amount of the reduction was calculated as a
function of model parameters in \cite{bfinhol, Carone:1995mx}.  The current
measurement of $R_b$ reported by the LEP Electroweak Working Group is
$R_b^{expt} = 0.21664 \pm 0.00068$.  This implies, at the 95\% c.l., that
$R_b$ lies no more than 0.26\% below the Standard Model value of 0.21583 .
Our figures show the contour $R_b - R_b^{SM} = -0.26\%$ in bold dots; the
allowed regions of parameter space lie above the contour.  For reference, the
contours at -0.1\% and -1.0\% are shown in light dots.

The predicted values of several other observables related to B physics trace
out curves in the model parameter space which are similar in shape to the
contours of constant $R_b$.  It is useful to compare them to get a sense of
the present and future constraints from heavy flavor physics.  First there is
the approximate limit from $B^0\bar{B}^0$ mixing (the ``B-line''), based on
requiring the estimated contributions from new physics in the model not to
exceed those from the Standard Model fields in the model. The constraint from
$R_b$ supersedes that imposed by the B-line, as illustrated in the figures.
Second, several authors have calculated the predicted rate of $b\to s\gamma$
in technicolor with scalars and related models \cite{gsw, mbsg, yumian,
  xiong} as a function of the model parameters.  The contour corresponding to
a 50\% reduction in the rate of $b\to s \gamma$ relative to the Standard
Model value is approximately contiguous with the B-line.  Recent measurements
of $b\to s \gamma$ from ALEPH \cite{xaleph}, BELLE\cite{xbelle}, and
CLEO\cite{xcleo} imply at 95\% c.l. that the maximum reduction relative to
the Standard Model rate\cite{smbsg} of 3.28 $\pm$ 0.33 $\times 10^{-4}$ is,
respectively, 78\%, 50\% and 48\%.  Hence, current experimental limits from
$b \to s\gamma$ are not significantly stronger than those from $B^0\bar{B}^0$
mixing, and are weaker than those from $R_b$.  More precise
measurements would have the power to test the model further.  Finally,
calculations of $B \to X_s \mu^+ \mu^-$ \cite{yumian}, $B \to X_s e^+e^-$
\cite{yumian}, and $B \to X_c \tau \bar{\nu}$ \cite{xiong} yield no currently
useful limits. Future experiments have the potential to make the first of
these a good probe of technicolor with scalars; the deviations from the
Standard Model values predicted for the other two are too small to be
visible.

\subsection{Neutral scalars}

The LEP Collaborations \cite{rbvalue} have placed 95\% c.l. lower
limit of $M_H \leq 113.5$ GeV on the mass of a neutral Higgs boson by
studying the process $Z^* \to Z H$ and assuming a Standard Model
coupling at the $ZZH$ vertex.  The $ZZ\sigma$ coupling in the
technicolor with scalars model is reduced relative to the standard
$ZZH$ coupling by a factor of $f' / v$, so that the LEP limit on
$m_\sigma$ differs, in principle, from that on $M_H$.  In practice,
however, in the region of parameter space still allowed by other
constraints, $f' / v \sim 1$ along the $m_\sigma = 114$ GeV contour.
This contour therefore serves as an approximate boundary to the
experimentally allowed region \cite{caronet}.  In limit [i], the bound
on $m_\sigma$ eliminates much of the parameter space for which
$\tilde{M}_\phi \leq 70$ GeV; in contrast, a just a few years ago
\cite{Carone:1995mx}, the limit on $m_\sigma$ was too weak to be
relevant.  In limit [ii], the bound on $m_\sigma$ excludes regions of
small $\tilde\lambda$ and obviates the theoretical restriction $h f'
\leq 4\pi f$.  Using the right-hand plots in Figures \ref{fig-limi} and \ref{fig-limii}, it is
straightforward to project how future experimental limits on
$m_\sigma$ will tend to constrain the model.

\subsection{Charged scalars}

The strongest limits on the charged physical scalars $\pi^\pm_p$ currently
come from LEP searches for the charged scalars characteristic of
two-higgs-doublet models.  The LEP experiments have obtained limits on the
charged scalar mass as a function of $\tan\beta$ and the branching ratio to
$\tau \nu$ final states (assuming all decays are to $\tau\nu$ or $cs$).  In
theories, like technicolor with scalars, where the charged scalar coupling to
fermions is of the pattern characteristic of type-I two-higgs models, the
branching fraction to $\tau\nu$ is predicted to be 1/3.  Hence, one can read
from figure 7 of ref. \cite{pdg} that the limit on $m_{\pi^\pm_p}$ is 78 GeV;
preliminary new data from LEP II \cite{chgd-higgs-79} pushes the lower bound to
79 GeV.  

The $m_{\pi^\pm_p} = 79$ GeV contour is shown in all of our figures for
reference, although the bounds on technicolor with scalars from data on
$m_\sigma$ and $R_b$ are currently stronger.  The contour $m_{\pi^\pm_p} =
m_t - m_b$ is also shown in each plot in order to indicate how stronger
bounds on charged scalar masses would tend to constrain the model.  Based on
the intersection of the current $m_\sigma$ and $R_b$ bounds, an experiment
sensitive to $m_{\pi^\pm_p} = 128$ (138) GeV would probe regions of limit [i]
(limit [ii]) parameter space beyond what is currently excluded.  If the lower
bound on $m_\sigma$ were to tighten to 133 (160) GeV in the future, then only
a search for charged scalars with $m_{\pi^\pm_p} \geq m_t - m_b$ would probe
regions of limit [i] (limit [ii]) beyond what limits from neutral scalars and
$R_b$ excluded.

The Tevatron experiments can search for light type-II charged scalars in top
quark decays.  While the Run I searches for charged scalars lacked the reach
of the LEP searches, that will change as Run II accumulates data.  For values
of $\tan\beta = f/f^\prime \lae 2$, the rate of $t \to H^\pm b$ is nearly
identical for type-I and type-II scalars; at higher $\tan\beta$, the rate for
type-I scalars drops off rapidly and the Tevatron limits do not directly
apply to technicolor with scalars.  D\O\ has set limits at low $\tan\beta$
based on the decay path $H^\pm \to cs$.  The value of $B(H^+\to c s)$ in
type-I models (2/3) matches the value in type-II models at $\tan\beta) = 1$.
Hence, one can read from figure 3 of \cite{dohigg} that the current limit from
D\O\ data is $m_{\pi^\pm_p} > 60$ GeV for $\tan\beta \leq 2$.  It is
projected that with $2 fb^{-1}$ of integrated luminosity the Run II
experiments will be sensitive to $\pi^\pm_p$ weighing up to 135 GeV
\cite{tampere}, a significant improvement over the LEP bounds at low
$\tan\beta$.

\section{Conclusions}

Technicolor with scalars remains a viable effective theory of dynamical
electroweak symmetry breaking and fermion mass generation.  Recent searches
for charged and neutral scalars and measurements of heavy flavor observables
such as $R_b$ have certainly reduced the extent of the allowed parameter
space.  However, the model is consistent with data for a wide range of
isosinglet scalar masses $m_\sigma$ and technifermion coupling to scalars
$h$.

In limit [i] of the model, where the scalar self-coupling $\lambda$ is small,
$m_\sigma$ is bounded from below by LEP searches for the higgs and from above
by a combination of the measured value of $R_b$ and the theoretical
consistency requirement $h f' = 4 \pi f$.  As shown in figure 1, 114 GeV
$\lae m_\sigma \lae$ 14 TeV.  In limit [ii], the constraint $h f' = 4 \pi f$
is superseded by the LEP limit $m_\sigma \gae$ 114 GeV.  Hence, larger values
of $h$ are allowed for a given $m_\sigma$ than in limit [i], as indicated in
figure 2, and the maximum allowed value of $m_\sigma$ is also somewhat
larger.  

Upcoming searches for charged and neutral scalar bosons will begin exploring
the lower allowed values in the $m_\sigma$ mass range.  At present, searches
for neutral scalars with masses above 114 GeV or charged scalars with masses
above 128 GeV (138 GeV) would give new information about limit [i] (limit
[ii]) of technicolor with scalars.  Complimenting this, new measurements of
$b \to s\gamma$ and $b \to s \mu^+ \mu^-$ will be sensitive even to the
heaviest allowed scalar masses.  If either branching ratio were
measured to be within a few percent of the standard model value, the
resulting exclusion curve in the $m_\sigma, h$ plane would run close to the
$\delta R_b / R_b = -0.1\%$ curves in figures 1 and 2 \cite{Carone:1995mx,
  yumian, xiong}, tending to reduce the largest allowed value of $m_\sigma$.

\newpage

\begin{center}
{\bf Acknowledgments}
\end{center}
EHS acknowledges the financial
support of an NSF POWRE Award. VH acknowledges the financial support
of a Radcliffe Research Partners Award. {\em This work was supported in
part by the National Science Foundation under grant PHY-0074274, and by the
Department of Energy under grant DE-FG02-91ER40676.}



\end{document}